\documentclass[aps,prl,preprint,superscriptaddress,longbibliography]{revtex4-1}

\usepackage{graphicx}
\usepackage{rotating}
\usepackage{color}
\usepackage{soul}
\usepackage[normalem]{ulem}
\usepackage{caption}
\usepackage{subcaption}
\usepackage{amsmath,amssymb}
\usepackage[english]{babel}
\usepackage{natbib}
\begin{document}

\title{Clustering of Janus Particles in Optical Potential Driven by Hydrodynamic Fluxes}

\author{S. Masoumeh Mousavi}
\affiliation{Soft Matter Lab, Department of Physics, Bilkent University, Ankara 06800, Turkey}

\author{Sabareesh K. P. Velu}
\affiliation{Soft Matter Lab, Department of Physics, Bilkent University, Ankara 06800, Turkey}

\author{Agnese Callegari}
\affiliation{Soft Matter Lab, Department of Physics, Bilkent University, Ankara 06800, Turkey}

\author{Luca Biancofiore}
\affiliation{Department of Mechanical Engineering, Bilkent University, Ankara 06800, Turkey}

\author{Giovanni Volpe}
\affiliation{Soft Matter Lab, Department of Physics, Bilkent University, Ankara 06800, Turkey}
\affiliation{Department of Physics, University of Gothenburg, SE-41296 Gothenburg, Sweden}

\maketitle

\section{Abstract}
Self-organisation is driven by the  interactions between the individual components of a system mediated by the environment, and is one of the most important strategies used by many biological systems to develop complex and functional structures. Furthermore, biologically-inspired self-organisation offers opportunities to develop the next generation of materials and devices for electronics, photonics and nanotechnology. In this work, we demonstrate experimentally that a system of Janus particles (silica microspheres half-coated with gold) aggregates into clusters in the presence of a Gaussian optical potential and disaggregates when the optical potential is switched off. We show that the underlying mechanism is the existence of a hydrodynamic flow induced by a temperature gradient generated by the light absorption at the metallic patches on the Janus particles. We also perform simulations, which agree well with the experiments and whose results permit us to clarify the underlying mechanism. The possibility of hydrodynamic-flux-induced reversible clustering may have applications in the fields of drug delivery, cargo transport, bioremediation and biopatterning. 

\section{1  \quad  Introduction}

Self-organisation entails the emergence of complex patterns and structures from relatively simple constituting building blocks \cite{zhang, van, jiang, walther, velu, mijalkov}. Phenomena such as flocking of birds and growth of bacterial colonies are examples of self-organisation in nature. Also artificial microscopic systems feature similar forms of organisation with the emergence of clusters, sometimes referred to as ``living crystals" \cite{palacci, buttinoni, gao, stenhammar, schmidt}.
In the past two decades, studies on self-organisation focused on systems made of complex colloids with anisotropic surface \cite{Perro, Pawar}, such as Janus particles \cite{walther,Gennes}. Depending on their surface material properties, Janus particles have been used in different fields for various applications such as self-assembly, microrheology and emulsion stabilisation \cite{jiang, walther}. 
Under certain conditions, Janus particles have the ability of self-propelling and behave as active Brownian particles \cite{howse, gangwal, volpe, buttinoni2012, illien}; these active Janus particles might be used in future biomedical nano-devices for diagnostics, drug delivery and microsurgery \cite{Wang, Baraban}.

Studies on clustering of Janus particles have been performed by Palacci et al. \cite{palacci}, who have shown the formation of living crystals in systems of light-activated Janus particles (Fe$_2$O$_3$-TPM) in hydrogen peroxide solution. 
Similarly, Buttinoni et al. \cite{buttinoni} demonstrated the clustering of light-activated Janus particles (carbon-SiO$_2$) in a water-lutidine binary mixture. 
Other research groups have shown self-assembly and controlled crystal formations in a mixed system of light-activated Janus particles and passive colloids \cite{gao, stenhammar}. 
In all these studies, a necessary ingredient for the clustering is the active nature of the particles. 
In systems of passive colloidal particles, crystallisation was observed at the bottom of an attractive optical potential \cite{pincce}, close to the hard boundary during electrophoretic deposition \cite{solomentsev}, and in the presence of an external temperature gradient \cite{weinert, di}.

Here, we investigate the behaviour of a system composed of Janus particles (silica microspheres half-coated with gold) close to a planar surface in the presence of an optical potential, and we experimentally demonstrate reversible clustering triggered by the presence of the optical field. Experimental results are compared and validated by numerical simulations, where the key ingredient for clustering is the presence of an attractive potential of hydrodynamic nature. Such results are confirmed also in mixtures of Janus particles and passive colloids (silica microspheres), where the hydrodynamic flux due to the Janus particles causes the clustering of the particles in the hybrid system and the formation of living crystals. As a further confirmation that the presence of Janus particles in the optical potential is crucial for the clustering, we show that a system with only non-Janus particles does not give rise to any clustering.

\section{2 \quad Experiments}

\begin{figure}[t!]
\centering
\includegraphics[width=12cm]{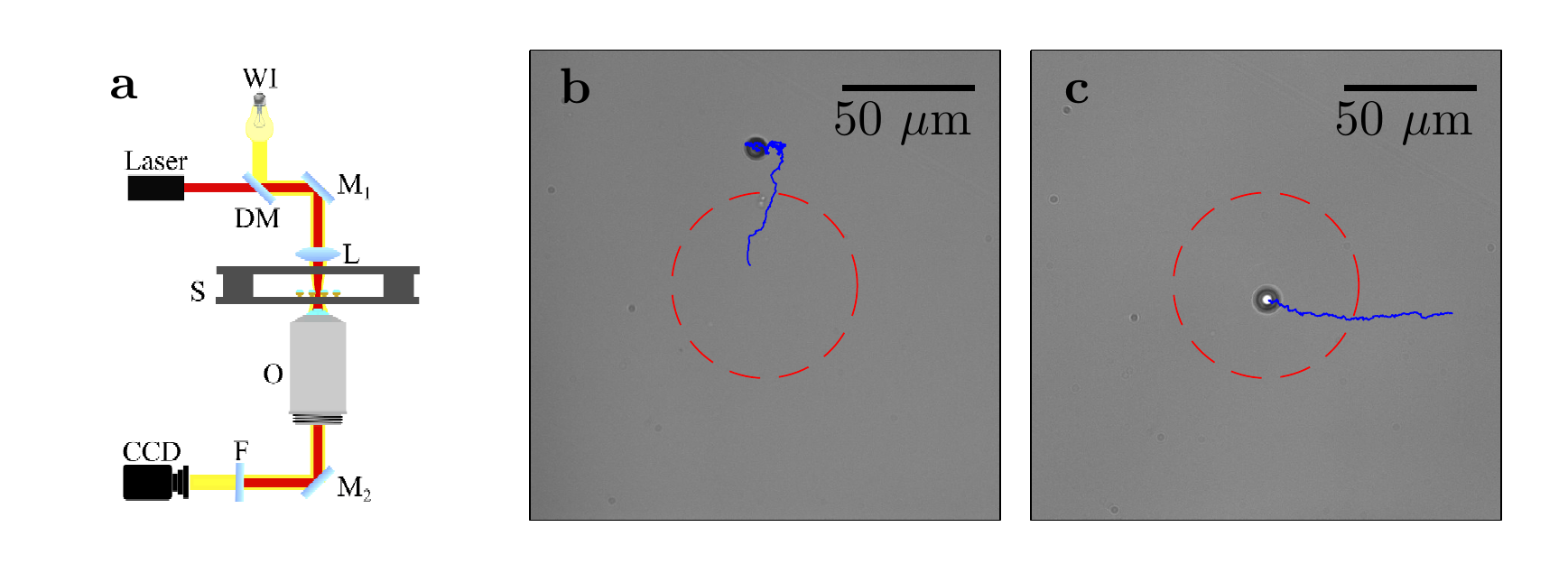}
\caption{Experimental setup and single particle behaviour. 
(a) Experimental setup to generate a Gaussian optical potential: WI, white light lamp; DM, dichroic mirror; M$_1$ and M$_2$, mirrors; L, convex lens; S, sample chamber; O, microscope objective; F, infrared filter; CCD, digital camera. 
(b) Typical trajectory of a single Janus particle in a broad Gaussian optical potential: it tends to move away from the region of maximum intensity towards the region of lower intensity. 
(c) Typical trajectory of a single silica particle in the same optical potential: it moves towards the region of maximum intensity.
The dashed red lines in (b) and (c) represent the region where the light intensity is more than the half of the maximum intensity.}
\label{Fig:1}
\end{figure}

The experiments are performed on a homemade inverted microscope, as schematically shown in Fig.~\ref{Fig:1}a. 
A laser beam (wavelength $\lambda = 976\,{\rm nm}$; power $P = 100\,{\rm mW}$) is focused by a convex lens (L, focal length $f = 50\,{\rm mm}$) onto the sample chamber (S) in order to generate a broad Gaussian  optical potential (beam waist $w_0 = 90\,{\rm \mu m}$) \cite{pesce}.
The height of the sample chamber is $200\,{\rm \mu m}$. 
The particles are tracked by digital video microscopy using the image projected by a microscope objective ($20\times$, ${\rm NA} = 0.50$) on a monochrome CCD camera with an acquisition rate of $5\,{\rm fps}$. 

In Fig.~\ref{Fig:1}(b), we show the typical motion of a Janus particle in the optical potential generated by the Gaussian laser beam. The Janus particle does not stay for a long time within the region of maximum intensity, but it is driven outwards by a combination of optical forces and optical torques: the presence of the reflecting thin gold layer results in an optical force directed towards the region of lower light intensity.
On the contrary, in Fig.~\ref{Fig:1}(c), we show that a silica particle is driven by the optical force towards the region of maximum intensity, as expected in the presence of optical forces \cite{jones}.

\begin{figure}[t!]
\centering
\includegraphics[width=16cm]{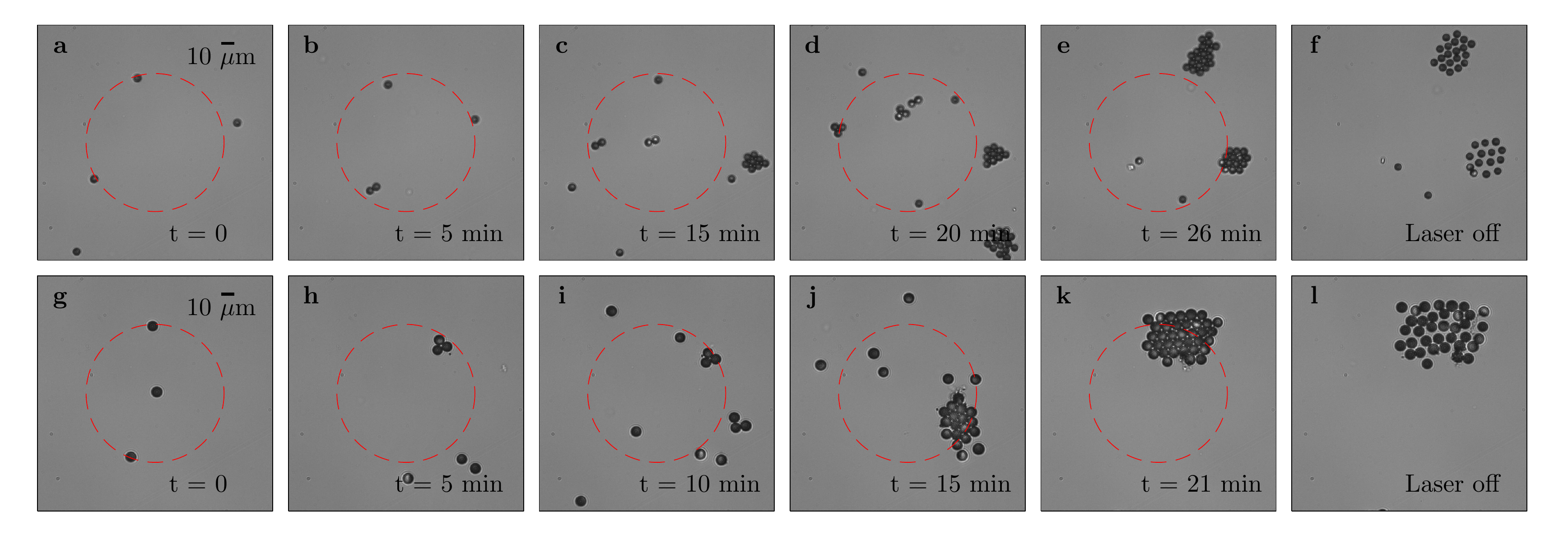}
\caption{ Experimental time sequences demonstrating the formation of clusters of Janus particles with diameter (a-f) $4.77\,{\rm \mu m}$ (see also Video 1) and (g-l) $6.73\,{\rm \mu m}$ (see also Video 2) in a Gaussian optical potential (beam waist $90\,{\rm \mu m}$, power $100\,{\rm mW}$). (f, l) As soon as the laser power is switched off, the clusters disassemble.}
\label{Fig:2}
\end{figure}

In Fig.~\ref{Fig:2}, we show the behaviour of multiple Janus particles in a  Gaussian optical potential. 
Figs.~\ref{Fig:2}(a-f) show a time sequence for a solution of Janus particles of $4.77\,{\rm \mu m}$ diameter, and Figs.~\ref{Fig:2}(g-l) a time sequence for a solution of Janus particles of $6.73\,{\rm \mu m}$ diameter.
In both cases, when the optical potential is turned on, the Janus particles cluster together and the centres of the clusters lay outside the centre of the optical potential. The process is slow at the beginning but accelerates as the size of the clusters increases. 
When the optical potential is switched off, the clusters immediately start to disassemble because of the attractive force between the particles disappears.

\begin{figure}[h!]
\centering
\includegraphics[width=16cm]{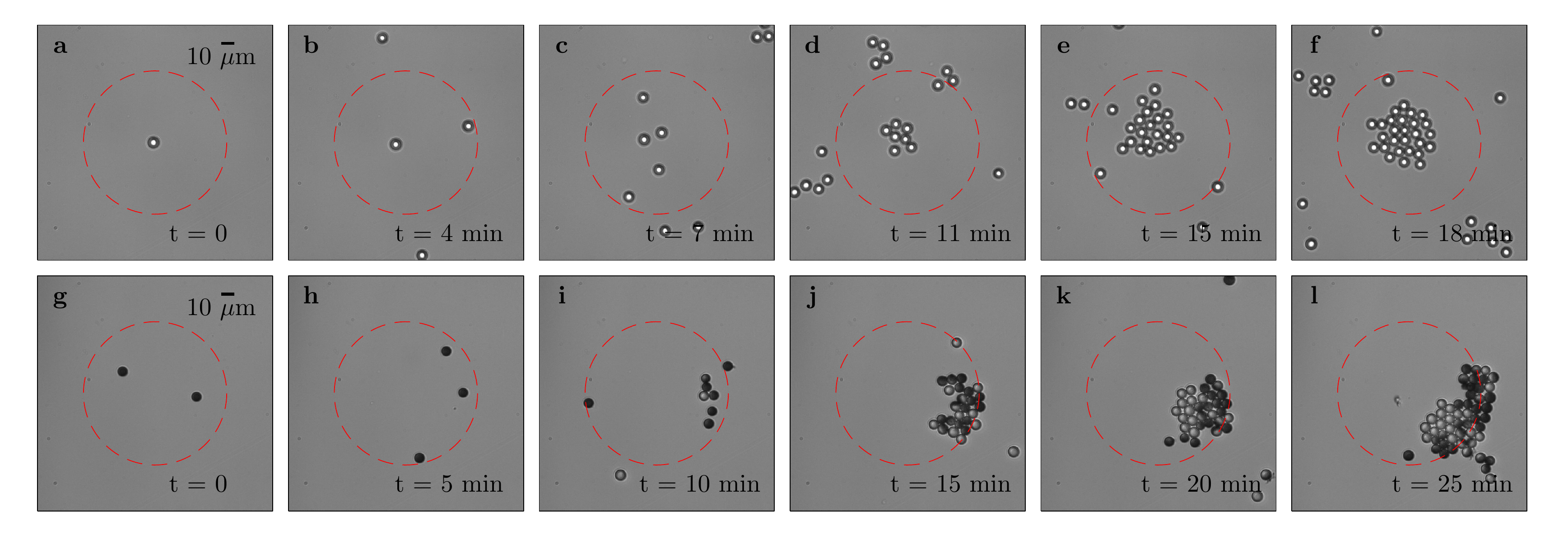}
\caption{Experimental time sequence of the dynamics (a-f) of silica particles (see also Video 3) and (g-l) of a mixture of Janus particles and silica particles (see also Video 4) (all particles have diameter $6.73\,{\rm \mu m}$) in a Gaussian optical potential (beam waist $90\,{\rm \mu m}$, power $100\,{\rm mW}$).
(a-f) The silica particles are pulled towards the centre of the optical potential, but do not form a close-packed colloidal crystal because of the absence of short-range attractive forces. 
(g-l) When also Janus particles are present, clusters form away from the centre of the optical potential, as in the cases with only Janus particles shown in Fig.~\ref{Fig:2}.}
\label{Fig:3}
\end{figure}

We do not observe this clustering behaviour in the case of a colloidal suspension composed only of silica particles with the same dilution as for the solution containing the Janus particles. 
In fact, when employing silica microspheres, aggregation and formation of a colloidal crystal are observed only with a significantly higher concentration (as observed, for example, in Ref.~\onlinecite{pincce}). In this case, the aggregation is not due to an effective attraction between the particles, but to the interplay between the optical forces pushing each particle towards the centre of the potential and the steric repulsion between the particles. The combined effect of these two interactions determines the formation of a regular close-packed lattice structure. However, this mechanism is ineffective at low concentrations, as shown in the time sequence presented in Figs.~\ref{Fig:3}(a-f): while the silica particles are attracted towards the centre of the potential, they do not form a cluster. 
If we add some Janus particles to this system of silica particles, the clustering of the particles away from the centre of the beam is recovered, as shown in Figs.~\ref{Fig:3}(g-l).

\section{3 \quad Model}

In order to understand the physical mechanism underlying the clustering of the Janus particles, we developed a numerical model of this system. In this model, we take into account the optical forces and torques acting on the Janus particles \cite{callegari, jones}, their Brownian motion \cite{fernandes, volpe2013simulation, jones}, and thermophoretic forces and torques \cite{bickel}. 

We model the optical forces and torques acting on a spherical Janus particle using the geometric optics approximation, because the size of the particle is significantly larger than the wavelength of the incident field and, in these conditions, the geometrical optics approximation reproduces well the features of the dynamics observed experimentally \cite{callegari, jones}. 
We model a Janus particle as a spherical dielectric microsphere plus a surface layer shaped as the hemispherical gold cap with a given thickness, mass density and refractive index. 
When the Janus particle is suspended in a solution and subject to an optical potential, there are three elements influencing its motion: 
(i) optical forces and torques due to the scattering of the light between media with different refractive indices \cite{callegari, jones};
(ii) Brownian forces due to the presence of a thermal noise \cite{fernandes, volpe2013simulation, jones}; 
and (iii) thermophoretic forces and torques, which are due to the partial light absorption by the golden cap determining a temperature gradient around the particle and, therefore, a self-propelled motion \cite{bickel}.
In addition, one should also take into account 
(iv) the combined effect of gravity and buoyancy, which keep the particles hovering just above the sample chamber bottom surface and;
and (v) the gravitational torque due to the inhomogeneity of the mass distribution of the Janus particle due to the gold coating, which, in the absence of any optical field, always results in a preferential downwards orientation of the golden cap.

In order to calculate the scattering and absorption of the golden cap, we use the thin film approximation for an absorbing layer on a transparent substrate \cite{heavens}. 
This permits us to obtain the reflectance, transmittance and absorbance of the metallic cap, and therefore to calculate the scattering of the light on the Janus particle. 
From the scattered rays, we obtain the optical force and torque according to the procedure in Refs.~\onlinecite{callegari, jones}.

In order to simulate the Brownian motion of a Janus particle, we have to take into account the asymmetry due to the presence of the metal cap, even though the shape of the Janus particle is accurately represented by a sphere. 
This entails that we need to take into account not only the translational motion, but also the rotational motion. 
Therefore, we use the $6 \times 6$ diffusion matrix, as in Ref.~\onlinecite{fernandes}. Furthermore, since the Janus particle are not in bulk but near a planar wall, we need to correct the translational and rotational diffusion for the effects of the close proximity to the boundary \cite{happel, lee}. 

The self-propelled motion originates from the presence of a local temperature gradient around the particle due to the light absorption by the metal-coated side of the Janus particle. A non-spherically symmetric temperature profile is induced around the particle due to the non-spherically symmetric shape of the absorbing layer. Such configuration induces a local force field tangential to the surface of the Janus particle.  This interfacial force leads to a slip velocity at the interface, i.e., a jump in the tangential fluid velocity component. This slip velocity drives the particle in the opposite direction along the temperature gradient axis \cite{bickel}, inducing the particle to self-propel. 
The velocity of this self-propulsion (i.e. the thermophoretic velocity) depends linearly on the temperature gradient, i.e. $v=- D_{\rm T} \nabla T_\parallel$, where $D_{\rm T} = \frac{a \kappa \gamma_T}{3\eta}$ is the thermophoretic mobility  (or thermal diffusion coefficient) \cite{weinert}. However, there is no certain law for the amplitude and sign of the thermophoretic mobility, which strongly depends on the microscopic nature of the particle-solvent interactions at the boundary layer of thickness $\lambda_{\rm D}$ (Debye length) \cite{dhont, ruckenstein, wurger, piazza}.  Depending on the sign of  $D_{\rm T}$,  the particle moves either towards the cold or the hot region \cite{weinert, piazza}.

In the case of two Janus particles close to a planar wall, a further effect of hydrodynamic nature has to be considered. This hydrodynamic effect creates an effective attraction among the particles. To explain this behaviour, one can use the same approach proposed for two immobile colloidal particles close to a wall \cite{weinert,morthomas,di}.
Indeed, a particle moves toward the horizontal bottom surface of the sample cell due to gravity, radiation force and interfacial driving force. Eventually, this particle would be fixed at a certain distance from the wall, due to the repulsive interaction with the wall and the viscous stress. The particle is then immobile \cite{weinert,morthomas,di} and the surrounding velocity field is squeezed by the boundary. Due to the temperature gradient surrounding the particles, the fluid continues to move along its surface. This creates a flow with a horizontal incoming radial component (parallel to the planar boundary) and outgoing vertical components, directed upwards from the wall. The thermophoretically-induced flow field  affects the motion of other neighbouring particles, so that a second nearby particle experiences an attractive hydrodynamic drag force toward the first particle.   
Following Ref.~\onlinecite{morthomas},  the radial component of the flow velocity at a horizontal distance $\rho$ from the centre of the immobile particle, $U_{\rho}$, is:
\begin{equation}\label{eq:radialflow}
U_{\rho} 
=
u_0 
\left[
6 \frac {\rho a h^3(\epsilon^2 q_3-p_1)}{\hat{r}_h^5}-60q_3 \frac{\rho a^3 h^3}{\hat{r}_h^7}
\right],
\end{equation}   
where $u_0=\frac{D_T \Delta T}{ \pi  a T_0}$, $\Delta T=\frac{P}{(2\pi+4)\kappa a} $, $P$ is the absorbed power by the gold cap with the total outward heat flow, $\kappa=3 \kappa_{\rm s} / (2\kappa_{\rm s}+\kappa_{\rm p})$ ($\kappa_{\rm s}$ and $\kappa_{\rm p}$ are the thermal conductivity of fluid and particle, respectively),  $p_1=1+\frac{9}{8} \epsilon$, $q_3=-1-\frac{3}{8} \epsilon$, $\epsilon=\frac{a}{h}$, $h$ is the distance of the centre of the particle with radius $a$ from the wall, and $\hat{r}_h=\sqrt{\rho^2+4h^2}$. 
From the flow velocity, $U_{\rho}$, one can obtain the effective hydrodynamic force on nearby particle as 
\begin{equation}
\label{brownian}
F=6 \pi \eta a \delta_{\rm w} U_{\rho},
\end{equation}   
where  $\delta_{\rm w}$ is a dimensionless factor which accounts for the effect of the presence of a planar wall on the effective friction coefficient in the direction parallel to the wall \cite{happel,di}. 
When this lateral flow is strong enough, the attractive hydrodynamic force can be larger than other repulsive contributions and than the thermal fluctuations, leading to a stable aggregation of the particles.
The strength of the attractive interaction can be regulated by the light intensity since the power absorbed by the gold cap determines the temperature gradient around the immobile particle and therefore the entity of the hydrodynamic lateral flow. 

Using this model we investigate in the next section the motion of a single and multiple Janus particles suspended in a water solution and in presence of a broad Gaussian optical potential.

\section{4 \quad Numerical results}

\begin{figure}[t!]
\centering
\includegraphics[width=12cm]{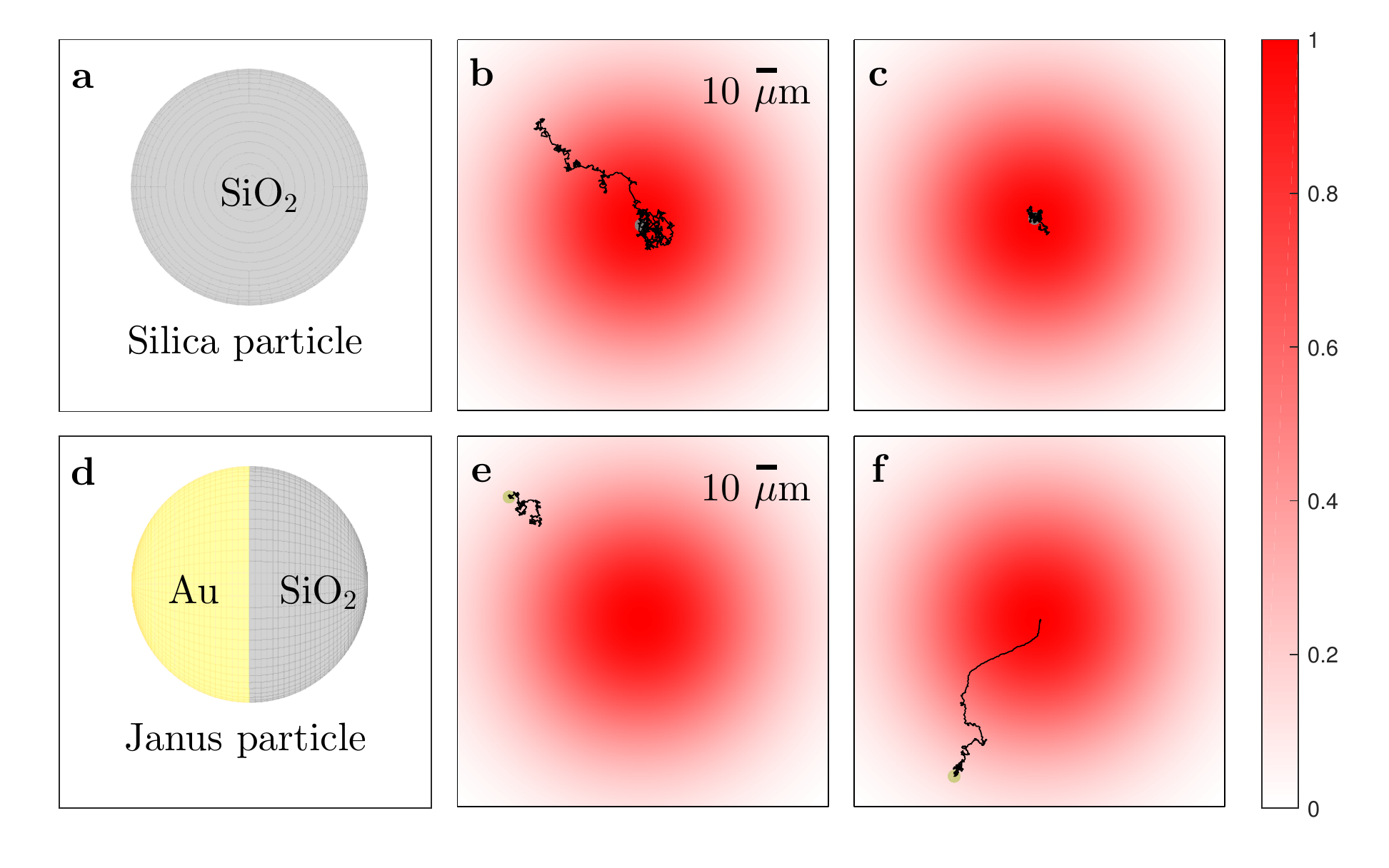}
\caption{Numerical simulation of the motion of (a) a silica particle and (d) a Janus particle with diameter $6.73\,{\rm \mu m}$ in the presence of Gaussian optical potential. 
(b) and (c) show the motion of the silica particle (black line) initially placed (small circle) at positions $(x_0,y_0) = (-50\,{\rm \mu m}, 50\,{\rm \mu m})$ (on the outer edge of the potential) and $(x_0,y_0) = (0,0)$ (centre of the potential). 
(e) and (f) show the same for the Janus particle.
The silica particle is attracted towards the centre of the Gaussian optical potential, while the Janus particle is attracted towards a circular region at a fixed radial distance from the potential centre. 
These results are in good agreement with the experiments shown in Figs.~\ref{Fig:1}(b) and \ref{Fig:1}(c).}
\label{Fig:4}
\end{figure}

In Fig.~\ref{Fig:4}, we show the behaviour of a silica particle (Fig.~\ref{Fig:4}(a-c)) and of a Janus particle (Fig.~\ref{Fig:4}(d-f)) in the presence of a Gaussian optical potential (radius $90\,{\rm \mu m}$, wavelength $976\,{\rm nm}$, power $100\,{\rm mW}$). The particles' trajectories are shown by the black lines. 
The particles are made of silica and have a diameter of $6.73\,{\rm \mu m}$, and the Janus particle is half-coated with a $60\,{\rm nm}$ gold layer (at the wavelength $976\,{\rm nm}$, the refractive indices of silica and gold film are $1.45$ and  $0.21-6.29i$, respectively \cite{johnson}). 
The silica particle (Fig.~\ref{Fig:4}(a)) moves toward the high intensity area and tends to remain in the region of higher optical intensity because of the presence of gradient optical forces \cite{ashkin, jones}, both when it is initially placed outside the potential (Fig.~\ref{Fig:4}(b)) and at its centre (Fig.~\ref{Fig:4}(c)). 
On the contrary, the Janus particle (Fig.~\ref{Fig:4}(d)) tends to move towards a circular region at a fixed distance from the centre of the potential, independently from whether it is initially placed outside the trapping potential (Fig.~\ref{Fig:4}(e)) or at its centre (Fig.~\ref{Fig:4}(f)).
These results are in agreement with the experiments shown in Figs.~\ref{Fig:1}(b) and \ref{Fig:1}(c).

\begin{figure}[t!]
\centering
\includegraphics[width=16cm]{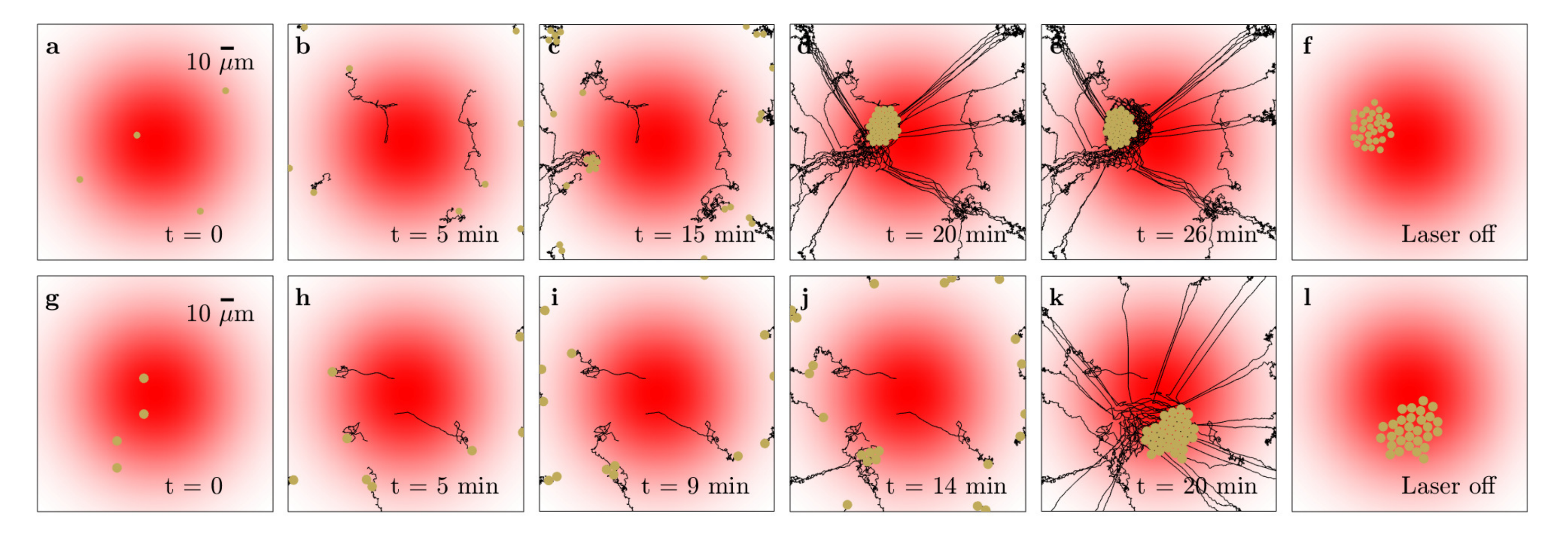}
\caption{Simulated time sequence demonstrating the formation of clusters of Janus particles for a system of 30 Janus particles with diameter (a-f) $4.77\,{\rm \mu m}$ (see also Video 1) and (g-l) $6.73\,{\rm \mu m}$ (see also Video 2) in a Gaussian optical potential. 
The black lines represent the trajectories of the particles.
The formation of the clusters, which are always located at a certain distance from the centre of the beam, is due to the hydrodynamic interaction among the particles. 
These simulations are consistent with the experimental results shown in Fig.~\ref{Fig:2}.}
 \label{Fig:5}
\end{figure}

In Fig.~\ref{Fig:5}, we show the time sequences corresponding to the clustering of Janus particles in a Gaussian optical potential.
As in the experiments shown in Fig.~\ref{Fig:2}, the diameter of the Janus particles is  $4.77\,{\rm \mu m}$ in Figs.~\ref{Fig:5}(a-f) and $6.73\,{\rm \mu m}$ in Figs.~\ref{Fig:5}(g-l). 
Both kinds of particles form clusters at a certain distance from the centre of the beam, which is in good agreement with the experiments shown in Fig.~\ref{Fig:2}. 
Furthermore, the clustering speed depends on the Janus particles size: larger particles aggregate more rapidly, as observed in experiments.

\begin{figure}[t!]
 \centering
 \includegraphics[width=16cm]{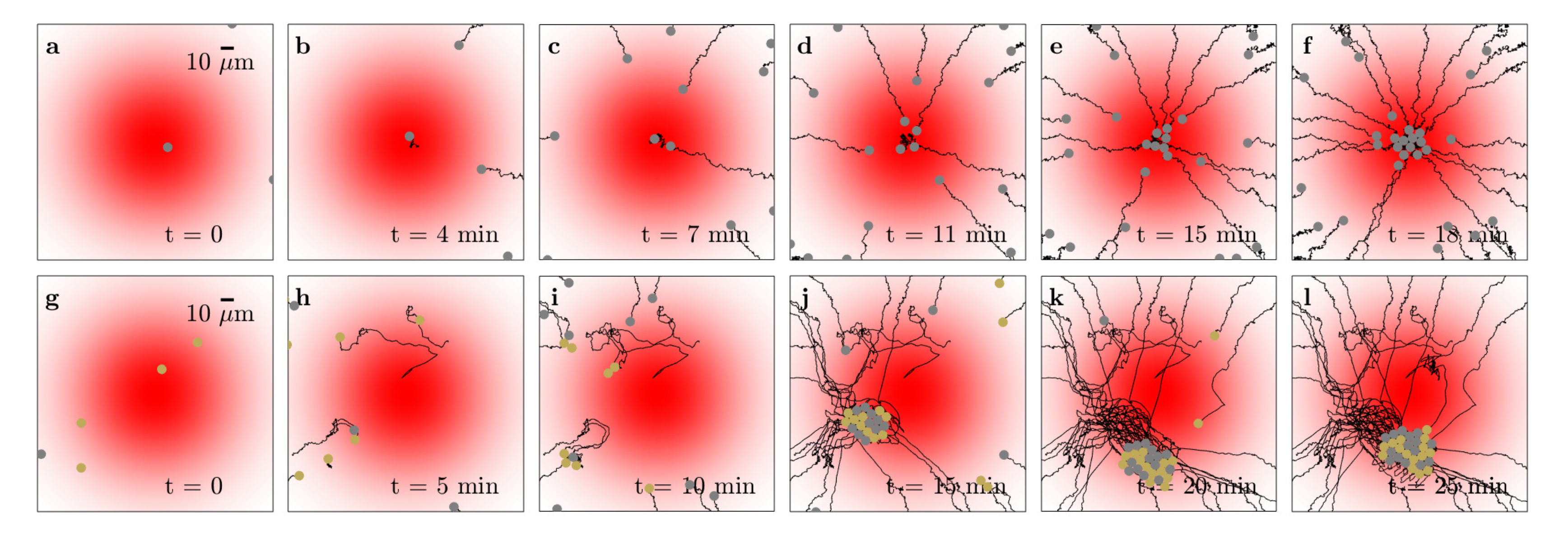}
\caption{Simulated time sequence of the dynamics (a-f) of silica particles (see also Video 3) and (g-l) of a mixture of Janus particles and silica particles (see also Video 4) in a Gaussian optical potential.
When alone, the silica particles do not feature any clustering behaviour apart from that due to the presence of optical forces driving them towards the centre of the potential.
Instead, in the presence of some Janus particles, the hydrodynamic flux generated by the Janus particles is strong enough to induce clustering of the particle mixture. The diameter of the particles is always $6.73\,{\rm \mu m}$. The black lines represent the trajectories of the particles.
These simulations are consistent with the experimental results shown in Fig.~\ref{Fig:3}.}
\label{Fig:6}
\end{figure}

In Fig.~\ref{Fig:6}(a-f), we show the simulated collective behaviour of a system of silica particles (diameter $6.73\,{\rm \mu m}$). In agreement with the experimental results shown in Figs.~\ref{Fig:3}(a-f), the silica particles go toward the centre of the Gaussian optical potential because of optical gradient forces, but do not form a colloidal crystal. 
In Fig.~\ref{Fig:6}(g-l), we show the simulated collective behaviour of a mixture of silica particles and Janus particles (diameter $6.73\,{\rm \mu m}$). Again in agreement with the experiments presented in Figs.~\ref{Fig:3}(g-l), the Janus particles generate a hydrodynamic flow that is sufficient to induce the clustering of all particles away from the optical potential center.

\section{5 \quad Conclusions}

We have shown with experiments and numerical simulations that the presence of Janus particles triggers the formation of clusters in a Gaussian optical potential. This is due to the presence of attractive hydrodynamic interactions among the particles. The presence of Janus particles is crucial for the cluster formation, since the attractive interaction is generated by the presence of a temperature gradient around the Janus particles: When a Janus particle is close to a boundary, this temperature gradient induces a hydrodynamic flow that drags other particles towards the Janus particle. In the absence of Janus particles, this hydrodynamic flow is absent and thus no clusters form. We have shown experimentally that the clustering process is reversible, since the cluster starts to disassemble as soon as the optical potential is switched off. 
Beyond their fundamental interest, the reported results are potentially relevant for various applications in the fields of self-assembly, targeted drug-delivery and bioremediation. For example, the possibility of forming clusters at a controllable distance from the minimum of a potential well offers a new route towards self-assembly near a target.
Future work will be devoted to understanding how the clustering behaviour can be controlled or altered by using more complex optical potentials.

\section{Acknowledgment}

SMM acknowledges a Tubitak 2216 fellowship and Tubitak project 114F207. 
SKPV acknowledges Tubitak projects 114F207 and 116F068. 
AC acknowledges Tubitak projects 115F401 and 116F111.

%


\begin{thebibliography}{41}
\providecommand{\natexlab}[1]{#1}
\providecommand{\url}[1]{\texttt{#1}}
\expandafter\ifx\csname urlstyle\endcsname\relax
  \providecommand{\doi}[1]{doi: #1}\else
  \providecommand{\doi}{doi: \begingroup \urlstyle{rm}\Url}\fi

\bibitem[Zhang and Glotzer(2004)]{zhang}
Z.~Zhang and S.~C. Glotzer.
\newblock Self-assembly of patchy particles.
\newblock \emph{Nanolett.}, 4\penalty0 (8):\penalty0 1407--1413, 2004.

\bibitem[van Blaaderen(2004)]{van}
A.~van Blaaderen.
\newblock Colloids under external control.
\newblock \emph{MRS Bull.}, 29\penalty0 (2):\penalty0 85--90, 2004.

\bibitem[Jiang et~al.(2010)Jiang, Chen, Tripathy, Luijten, Schweizer, and
  Granick]{jiang}
S.~Jiang, Q.~Chen, M.~Tripathy, E.~Luijten, K.~S. Schweizer, and S.~Granick.
\newblock Janus particle synthesis and assembly.
\newblock \emph{Adv. Mat.}, 22\penalty0 (10):\penalty0 1060--1071, 2010.

\bibitem[Walther and M{\"u}ller(2013)]{walther}
A.~Walther and A.~H.~E. M{\"u}ller.
\newblock Janus particles: Synthesis, self-assembly, physical properties, and
  applications.
\newblock \emph{Chem. Rev.}, 113\penalty0 (7):\penalty0 5194--5261, 2013.

\bibitem[Velu et~al.(2013)Velu, Yan, Tseng, Wong, Bassani, and Terech]{velu}
S.~K.~P. Velu, M.~Yan, K.-P. Tseng, K.-T. Wong, D.~M Bassani, and P.~Terech.
\newblock Spontaneous formation of artificial vesicles in organic media through
  hydrogen-bonding interactions.
\newblock \emph{Macromolecules}, 46\penalty0 (4):\penalty0 1591--1598, 2013.

\bibitem[Mijalkov et~al.(2016)Mijalkov, McDaniel, Wehr, and Volpe]{mijalkov}
M.~Mijalkov, A.~McDaniel, J.~Wehr, and G.~Volpe.
\newblock Engineering sensorial delay to control phototaxis and emergent
  collective behaviors.
\newblock \emph{Phys. Rev. X}, 6\penalty0 (1):\penalty0 011008, 2016.

\bibitem[Palacci et~al.(2013)Palacci, Sacanna, Steinberg, Pine, and
  Chaikin]{palacci}
J.~Palacci, S.~Sacanna, A.~P. Steinberg, D.~J. Pine, and P.~M. Chaikin.
\newblock Living crystals of light-activated colloidal surfers.
\newblock \emph{Science}, 339\penalty0 (6122):\penalty0 936--940, 2013.

\bibitem[Buttinoni et~al.(2013)Buttinoni, Bialk{\'e}, K{\"u}mmel, L{\"o}wen,
  Bechinger, and Speck]{buttinoni}
Ivo Buttinoni, J.~Bialk{\'e}, F.~K{\"u}mmel, H.~L{\"o}wen, C.~Bechinger, and
  T.~Speck.
\newblock Dynamical clustering and phase separation in suspensions of
  self-propelled colloidal particles.
\newblock \emph{Phys. Rev. Lett.}, 110\penalty0 (23):\penalty0 238301, 2013.

\bibitem[Gao et~al.(2013)Gao, Pei, Feng, Hennessy, and Wang]{gao}
W.~Gao, A.~Pei, X.~Feng, C.~Hennessy, and J.~Wang.
\newblock Organized self-assembly of janus micromotors with hydrophobic
  hemispheres.
\newblock \emph{J. Am. Chem. Soc.}, 135\penalty0 (3):\penalty0 998--1001, 2013.

\bibitem[Stenhammar et~al.(2015)Stenhammar, Wittkowski, Marenduzzo, and
  Cates]{stenhammar}
J.~Stenhammar, R.~Wittkowski, D.~Marenduzzo, and M.~E. Cates.
\newblock Activity-induced phase separation and self-assembly in mixtures of
  active and passive particles.
\newblock \emph{Phys. Rev. Lett.}, 114\penalty0 (1):\penalty0 018301, 2015.

\bibitem[Schmidt et~al.(2018)Schmidt, Liebchen, L{\"o}wen, and Volpe]{schmidt}
F.~Schmidt, B.~Liebchen, H.~L{\"o}wen, and G.~Volpe.
\newblock Light-controlled assembly of active colloidal molecules.
\newblock \emph{arXiv}, page 1801.06868, 2018.

\bibitem[Perro et~al.(2005)Perro, Reculusa, Ravaine, Bourgeat-Lami, and
  Duguet]{Perro}
A.~Perro, S.~Reculusa, S.~Ravaine, E.~Bourgeat-Lami, and E.~Duguet.
\newblock Design and synthesis of janus micro-and nanoparticles.
\newblock \emph{J. Mat. Chem.}, 15\penalty0 (35-36):\penalty0 3745--3760, 2005.

\bibitem[Pawar and Kretzschmar(2010)]{Pawar}
A.~B. Pawar and I.~Kretzschmar.
\newblock Fabrication, assembly, and application of patchy particles.
\newblock \emph{Macromol. Rap. Commun.}, 31\penalty0 (2):\penalty0 150--168,
  2010.

\bibitem[de~Gennes(1992)]{Gennes}
P.~G. de~Gennes.
\newblock Soft matter.
\newblock \emph{Science}, 256\penalty0 (5056):\penalty0 495--497, 1992.

\bibitem[Howse et~al.(2007)Howse, Jones, Ryan, Gough, Vafabakhsh, and
  Golestanian]{howse}
J.~R. Howse, R.~A.~L. Jones, A.~J. Ryan, T.~Gough, R.~Vafabakhsh, and
  R.~Golestanian.
\newblock Self-motile colloidal particles: From directed propulsion to random
  walk.
\newblock \emph{Phys. Rev. Lett.}, 99\penalty0 (4):\penalty0 048102, 2007.

\bibitem[Gangwal et~al.(2008)Gangwal, Cayre, Bazant, and Velev]{gangwal}
S.~Gangwal, O.~J. Cayre, M.~Z. Bazant, and O.~D. Velev.
\newblock Induced-charge electrophoresis of metallodielectric particles.
\newblock \emph{Phys. Rev. Lett.}, 100\penalty0 (5):\penalty0 058302, 2008.

\bibitem[Volpe et~al.(2011)Volpe, Buttinoni, Vogt, K{\"u}mmerer, and
  Bechinger]{volpe}
G.~Volpe, I.~Buttinoni, D.~Vogt, H.-J. K{\"u}mmerer, and C.~Bechinger.
\newblock Microswimmers in patterned environments.
\newblock \emph{Soft Matter}, 7\penalty0 (19):\penalty0 8810--8815, 2011.

\bibitem[Buttinoni et~al.(2012)Buttinoni, Volpe, K\"{u}mmel, Volpe, and
  Bechinger]{buttinoni2012}
I.~Buttinoni, G.~Volpe, F.~K\"{u}mmel, G.~Volpe, and C.~Bechinger.
\newblock Active {B}rownian motion tunable by light.
\newblock \emph{J. Phys.: Condens. Matter}, 24:\penalty0 284129, 2012.

\bibitem[Illien et~al.(2017)Illien, Golestanian, and Sen]{illien}
P.~Illien, R.~Golestanian, and A.~Sen.
\newblock `{F}uelled' motion: Phoretic motility and collective behaviour of
  active colloids.
\newblock \emph{Chem. Soc. Rev.}, 46:\penalty0 5508, 2017.

\bibitem[Wang and Gao(2012)]{Wang}
J.~Wang and W.~Gao.
\newblock Nano/microscale motors: biomedical opportunities and challenges.
\newblock \emph{ACS Nano}, 6\penalty0 (7):\penalty0 5745--5751, 2012.

\bibitem[Baraban et~al.(2012)Baraban, Makarov, Streubel, M{\"o}nch, Grimm,
  Sanchez, and Schmidt]{Baraban}
L.~Baraban, D.~Makarov, R.~Streubel, I.~M{\"o}nch, D.~Grimm, S.~Sanchez, and
  O.~G. Schmidt.
\newblock Catalytic janus motors on microfluidic chip: deterministic motion for
  targeted cargo delivery.
\newblock \emph{ACS Nano}, 6\penalty0 (4):\penalty0 3383--3389, 2012.

\bibitem[Pin{\c{c}}e et~al.(2016)Pin{\c{c}}e, Velu, Callegari, Elahi, Gigan,
  Volpe, and Volpe]{pincce}
E.~Pin{\c{c}}e, S.~K.~P. Velu, A.~Callegari, P.~Elahi, S.~Gigan, G.~Volpe, and
  G.~Volpe.
\newblock Disorder-mediated crowd control in an active matter system.
\newblock \emph{Nature Commun.}, 7:\penalty0 10907, 2016.

\bibitem[Solomentsev et~al.(1997)Solomentsev, B{\"o}hmer, and
  Anderson]{solomentsev}
Y.~Solomentsev, M.~B{\"o}hmer, and J.~L. Anderson.
\newblock Particle clustering and pattern formation during electrophoretic
  deposition: A hydrodynamic model.
\newblock \emph{Langmuir}, 13\penalty0 (23):\penalty0 6058--6068, 1997.

\bibitem[Weinert and Braun(2008)]{weinert}
F.~M. Weinert and D.~Braun.
\newblock Observation of slip flow in thermophoresis.
\newblock \emph{Phys. Rev. Lett.}, 101\penalty0 (16):\penalty0 168301, 2008.

\bibitem[Di~Leonardo et~al.(2009)Di~Leonardo, Ianni, and Ruocco]{di}
R.~Di~Leonardo, F.~Ianni, and G.~Ruocco.
\newblock Colloidal attraction induced by a temperature gradient.
\newblock \emph{Langmuir}, 25\penalty0 (8):\penalty0 4247--4250, 2009.

\bibitem[Pesce et~al.(2015)Pesce, Volpe, Marag\`{o}, Jones, Gigan, Sasso, and
  Volpe]{pesce}
G.~Pesce, G.~Volpe, O.~M. Marag\`{o}, P.~H. Jones, S.~Gigan, A.~Sasso, and
  G.~Volpe.
\newblock Step-by-step guide to the realization of advanced optical tweezers.
\newblock \emph{J. Opt. Soc. Am. B}, 32:\penalty0 B84--B98, 2015.

\bibitem[Jones et~al.(2015)Jones, Marag{\`o}, and Volpe]{jones}
P.~H. Jones, O.~M. Marag{\`o}, and G.~Volpe.
\newblock \emph{Optical tweezers: Principles and applications}.
\newblock Cambridge University Press, 2015.

\bibitem[Callegari et~al.(2015)Callegari, Mijalkov, G{\"o}k{\"o}z, and
  Volpe]{callegari}
A.~Callegari, M.~Mijalkov, A.~B. G{\"o}k{\"o}z, and G.~Volpe.
\newblock Computational toolbox for optical tweezers in geometrical optics.
\newblock \emph{J. Opt. Soc. Am. B}, 32\penalty0 (5):\penalty0 B11--B19, 2015.

\bibitem[Fernandes and de~la Torre(2002)]{fernandes}
M.~X. Fernandes and J.~G. de~la Torre.
\newblock Brownian dynamics simulation of rigid particles of arbitrary shape in
  external fields.
\newblock \emph{Biophys. J.}, 83\penalty0 (6):\penalty0 3039--3048, 2002.

\bibitem[Volpe and Volpe(2013)]{volpe2013simulation}
G.~Volpe and G.~Volpe.
\newblock Simulation of a {B}rownian particle in an optical trap.
\newblock \emph{Am. J. Phys.}, 81\penalty0 (3):\penalty0 224--230, 2013.

\bibitem[Bickel et~al.(2013)Bickel, Majee, and W{\"u}rger]{bickel}
T.~Bickel, A.~Majee, and A.~W{\"u}rger.
\newblock Flow pattern in the vicinity of self-propelling hot janus particles.
\newblock \emph{Phys. Rev. E}, 88\penalty0 (1):\penalty0 012301, 2013.

\bibitem[Heavens(1991)]{heavens}
O.~S. Heavens.
\newblock \emph{Optical properties of thin solid films}.
\newblock Courier Corporation, 1991.

\bibitem[Happel and Brenner(2012)]{happel}
J.~Happel and H.~Brenner.
\newblock \emph{Low {R}eynolds number hydrodynamics: with special applications
  to particulate media}, volume~1.
\newblock Springer Science \& Business Media, 2012.

\bibitem[Lee et~al.(1979)Lee, Chadwick, and Leal]{lee}
S.~H. Lee, R.~S. Chadwick, and L.~G. Leal.
\newblock Motion of a sphere in the presence of a plane interface. part 1. an
  approximate solution by generalization of the method of {L}orentz.
\newblock \emph{J. Fluid Mech.}, 93\penalty0 (4):\penalty0 705--726, 1979.

\bibitem[Dhont et~al.(2007)Dhont, Wiegand, Duhr, and Braun]{dhont}
J.~K.~G. Dhont, S.~Wiegand, S.~Duhr, and D.~Braun.
\newblock Thermodiffusion of charged colloids: Single-particle diffusion.
\newblock \emph{Langmuir}, 23\penalty0 (4):\penalty0 1674--1683, 2007.

\bibitem[Ruckenstein(1981)]{ruckenstein}
E.~Ruckenstein.
\newblock Can phoretic motions be treated as interfacial tension gradient
  driven phenomena?
\newblock \emph{J. Colloid Interfac. Sci.}, 83\penalty0 (1):\penalty0 77--81,
  1981.

\bibitem[W{\"u}rger(2007)]{wurger}
A.~W{\"u}rger.
\newblock Thermophoresis in colloidal suspensions driven by {M}arangoni forces.
\newblock \emph{Phys. Rev. Lett.}, 98\penalty0 (13):\penalty0 138301, 2007.

\bibitem[Piazza(2008)]{piazza}
R.~Piazza.
\newblock Thermophoresis: moving particles with thermal gradients.
\newblock \emph{Soft Matter}, 4\penalty0 (9):\penalty0 1740--1744, 2008.

\bibitem[Morthomas and W{\"u}rger(2010)]{morthomas}
J.~Morthomas and A.~W{\"u}rger.
\newblock Hydrodynamic attraction of immobile particles due to interfacial
  forces.
\newblock \emph{Phys. Rev. E}, 81\penalty0 (5):\penalty0 051405, 2010.

\bibitem[Johnson. and Christy(1972)]{johnson}
Peter~B. Johnson. and R.-W. Christy.
\newblock Optical constants of the noble metals.
\newblock \emph{Phys. Rev. B}, 6\penalty0 (12):\penalty0 4370, 1972.

\bibitem[Ashkin et~al.(1986)Ashkin, Dziedzic, Bjorkholm, and Chu]{ashkin}
A.~Ashkin, J.~M. Dziedzic, J.~E. Bjorkholm, and S.~Chu.
\newblock Observation of a single-beam gradient force optical trap for
  dielectric particles.
\newblock \emph{Opt. Lett.}, 11\penalty0 (5):\penalty0 288--290, 1986.

\end{thebibliography}

\end{document}